\documentclass[aps,nofootinbib]{revtex4}

\usepackage[normalem]{ulem}

\usepackage{amsmath}
\usepackage{amsfonts,amssymb,amsthm, bbm,braket}
\usepackage{graphicx}   
\usepackage{subfigure}  
\usepackage{amsbsy} 
\usepackage[bold]{hhtensor} 
\usepackage{natbib}
\usepackage{algpseudocode}

\usepackage{multirow}

\newcommand\Tr{\mathrm{Tr}}

\usepackage{hyperref} 
\usepackage{algorithm} 

\begin{document}

\title{High posterior density ellipsoids of quantum states}

\author{Christopher Ferrie}
\affiliation{
Center for Quantum Information and Control,
University of New Mexico,
Albuquerque, New Mexico, 87131-0001}

\date{\today}


\begin{abstract}
Regions of quantum states generalize the classical notion of error bars.  High posterior density (HPD) credible regions are the most powerful of region estimators.  However, they are intractably hard to construct  in general.  This paper reports on a numerical approximation to HPD regions for the purpose of testing a much more computationally and conceptually convenient class of regions: posterior covariance ellipsoids (PCEs).  The PCEs are defined via the covariance matrix of the posterior probability distribution of states.   Here it is shown that PCEs are near optimal for the example of Pauli measurements on multiple qubits.  Moreover, the algorithm is capable of producing accurate PCE regions even when there is uncertainty in the model.
\end{abstract}


\maketitle

\section{Introduction}

Characterizing a physical system at the most fundamental level requires specifying its quantum mechanical state.  Arriving at such a description from measured data is called quantum tomography and the output of such a process is often a single point in the space of allowed parameters.  In theory, by considering an infinite amount of data, a unique state can be identified.  In practice, however, only a finite amount of data can be obtained.  In such cases, it is impossible for a single reported state to coincide with the true state.  In classical data fitting, \emph{error bars} give a measure of the \emph{accuracy} of the estimate.  In the quantum state tomography setting, \emph{regions} generalize this concept \cite{BlumeKohout2010Optimal}.  A region of quantum states should colloquially be understood to contain the true state with high probability (with the exact interpretation depending subtly on how the region is constructed).

Although this idea is quite simple, formalizing the concept of region estimation is not straightforward and there exists many competing alternatives, each with its own set of advantages and drawbacks.  For example, \emph{bootstrap resampling} is a common technique to produce error bars in tomographic experiments.  However, as in  \cite{Home2009Complete} for example, bootstrapping has so far been exclusively used to calculate statistics on quantities derived from state estimates, such as fidelity to some target state.  Bootstrapping is conceptually simple and easy to implement.  However, the errors bootstrapping estimate come with no guarantees and it can grossly underestimate errors for estimators which produce states near the boundary \cite{BlumeKohout2012Robust}.  The Fisher information (matrix) is most often used, via the Cramer-Rao inequality, to lower bound the variance of unbiased estimators.  In this sense, it gives the errors one would expect in the asymptotic limit, provided an efficient estimator is used.  In terms of regions, the Fisher information matrix is also asymptotically the inverse of the covariance matrix of the posterior distribution of parameters, which in turn defines an error ellipse \cite{Rehacek2008Tomography}.  For most problems, however, even computing the Fisher information numerically is an intractable problem.  In some estimation strategies, such as compressed sensing \cite{Flammia2012Quantum} (also \cite{Sugiyama2011Error,Sugiyama2013Precisionguaranteed}), the estimate of the state comes with a \emph{certificate}.  That is, an estimated state is provided along with an upper bound on the distance to the true state.  This implicitly defines a ball in state space centered on the estimated state.  However, the statistical meaning of this ball is not clear nor does a ball provide information about correlated errors.  Confidence regions \cite{Christandl2011Reliable,BlumeKohout2012Robust} (and comparable constructions \cite{Shang2013Optimal}) are the most stringently defined regions from a statistical perspective.  These regions would be ideal if they admitted a method of construction which is computationally tractable.

There is one overarching theme to notice here: trade-offs.  On one end of the spectrum is conceptual simplicity, ease of implementation and computational tractability; on the other is statistical rigor, precision, accuracy and optimality.  Here we take the approach of constructing statistically rigorous region estimators via a numerical algorithm which possess tunable parameters to control the trade-off between optimality and computational efficiency.  The regions we construct here are approximations to \emph{high posterior density credible regions} and are in some sense the Bayesian analogs of confidence regions.  To aid in the descriptive simplicity of the regions, we use \emph{ellipsoids} which are well understood and easy to conceptualize geometrical objects.  Moreover, ellipsoids provide a useful summary of how the state parameters are correlated. 

The numerical algorithm we use is a Monte Carlo approximation to Bayesian inference and has been used to in the tomographic estimation of one and two qubit states \cite{Huszar2012Adaptive} as well as in the continuous measurement of a qubit \cite{Chase2009Singleshot}.  It has also been recently used to construct point \emph{and} region estimates of Hamiltonian parameters in \cite{Granade2012Robust,Wiebe2013Hamiltonian}.  In particular, the region used was the ellipse defined by the posterior covariance matrix.  Here we show that the same method can be applied to quantum states and, more importantly, such regions approximate high posterior density credible regions.

One of the major advantages to this approach is that it naturally accommodates the possibility of unknown errors in modeling. For example, we might assume that the source of quantum states is fixed when it is not; or, we might assume that the measurements are known exactly when they are not.  Previous analysis of errors in quantum state estimation have focused on assessing its effect \cite{Rosset2012Imperfect}; \emph{detecting} its presence \cite{Moroder2013Certifying, Langford2013Errors}; and, most recently, selecting the best model for it \cite{vanEnk2013When,Schwarz2013Error}.  Here we demonstrate that the our approach can estimate and construct regions for both quantum mechanical state and error model parameters \emph{simultaneously}. That is, our algorithm produces a region in the space defined by all parameters.

This work is outlined as follows.  In section \ref{sec:bayes} we overview the precise problem and theoretical solution.  In section \ref{sec:smc} we give the numerical algorithm which constructs the regions.  In section \ref{sec:ex} we describe the examples used to test the method and in section \ref{sec:results} the results of the numerical experiments are presented.  Finally, in section \ref{sec:end} we conclude discussion.

\section{Bayesian Learning and Credible Regions}\label{sec:bayes}

Each quantum mechanical problem specification produces a probability distribution $\Pr(d_k|\vec{x};c_k)$, where $d_k$ is the data obtained and $c_k$ are the experimental designs (or \emph{controls}) chosen for measurement $k$, and where $\vec{x}$ is a vector parameterizing the system of interest.

Suppose we have performed experiments with control settings $C:=\{c_1,c_2,\ldots,c_N\}$ and obtained data $D:=\{d_1,d_2,\ldots,d_N\}$.  The model specifies the likelihood function
\begin{equation}
\Pr(D|\vec{x};C) = \prod_{k=1}^N \Pr(d_k|\vec{x};c_k).
\end{equation}
However, we are ultimately interested in $\Pr(\vec{x}|D;C)$, the probability distribution of the model parameters $\vec{x}$ given the experimental data.  We achieve this using use Bayes' rule:
\begin{equation}
\Pr(\vec{x}|D;C)=\frac{\Pr(D|\vec{x};C)\Pr(\vec{x})}{\Pr(D|C)},
\end{equation}
where $\Pr(\vec{x})$ is the \emph{prior}, which encodes any \emph{a priori} knowledge of the model parameters.  The final term $\Pr(D|C)$ can simply be thought as a normalization factor.  Since each measurement is statistically independent given $\vec{x}$, the processing of the data can be done on or off-line.  That is, we can sequentially update the probability distribution as the data arrive or post-process it afterward.   

In many scenarios the mean of the posterior distribution\footnote{We use the following notation for expectation values with respect to the distribution of the random variable, say, $X$: $\mathbb E_X[f(X)]=\int f(X)\Pr(X) dx$.}
\begin{equation}
\hat{\vec{x}}(D;C) = \mathbb{E}_{\vec{x}|D;C}[\vec{x}]
\end{equation}
is the optimal \emph{point} estimator \cite{BlumeKohout2006Accurate}.  But we desire more than point estimates.  In this scenario, the most powerful region estimators are \emph{high posterior density} (HPD) regions.  Let $\mathbbm{1}_{X}$ be the indicator function of the set $X$.  That is,
\begin{equation}
\mathbbm{1}_{X}(\vec{x}) = \begin{cases} 1 &{\rm if }\;\; \vec{x}\in X \\0 &{\rm if }\;\; \vec{x}\not\in\ X 
\end{cases}.
\end{equation}
Then, the probability that $\vec{x}$ lies in some set $X$ is
\begin{equation}
\Pr(\vec{x}\in X|D;C) =  \mathbb{E}_{\vec{x}|D;C}[\mathbbm{1}_{X}(\vec{x})].
\end{equation}
The set $X$ is an $\alpha$-credible region if
\begin{equation}
\Pr(\vec{x}\in X|D;C) \geq 1-\alpha.
\end{equation}
That is, a set is $\alpha$-credible region if no more than $\alpha$ probability mass exists outside the region or, equivalently, at least $1-\alpha$ probability mass is contained within the region.  The set $X$ is an HPD $\alpha$-credible region if
\begin{equation}\label{eq:HPD}
X = \{\vec{x}:\Pr(\vec{x}|D;C)\geq y\},
\end{equation}
where $y$ is the largest number such that $X$ is a $\alpha$-credible interval.  Under natural notions of ``size'', HPD $\alpha$-credible are the \emph{smallest} (that is, most powerful) $\alpha$-credible regions.

 Being optimal, HPD regions are the solutions to computationally intractable optimization problems---at least if we require a deterministic algorithm.  In this work, we will explore a Monte Carlo algorithm to numerically approximate HPD regions.  In particular, we use \emph{posterior covariance ellipsoids} (PCEs).  These are defined via the mean and covariance matrix of the posterior probability distribution as follows\footnote{The covariance matrix of a vector random variable $X$ is denoted ${\rm Cov}_X[X] = \mathbb E_X[(X-\mathbb E[X])(X-\mathbb E[X])^{\rm T}]$.}:
\begin{equation}\label{eq:PCE}
X_{\rm PCE} = \{\vec{x}: (\vec{x}-\mathbb{E}_{\vec{x}|D;C}[\vec{x}])^{\rm T}{\rm Cov}_{\vec{x}|D;C}[\vec{x}]^{-1}(\vec{x}-\mathbb{E}_{\vec{x}|D;C}[\vec{x}]) \leq Z_\alpha^2\}
\end{equation}
where $Z_\alpha^2$ is the $\alpha$-quantile of the $\chi_d^2$ distribution \cite{Robert2007The}, the values of which are readily available in countless tables.  Note that this is simply the covariance ellipse under a Gaussian approximation to the posterior $\vec{x}|D;C \sim \mathcal N(\mathbb{E}_{\vec{x}|D;C}[\vec{x}],{\rm Cov}_{\vec{x}|D;C}[\vec{x}])$.
In addition to being HPD credible regions in the asymptotic limit, PCEs are computationally tractable and, even for modest numbers of experiments, they are remarkably close in size and coverage probability to true HPD regions.  The remainder of this work is devoted to detailing the algorithm and demonstrating the above claims via simulated experiments on qubits.  

\section{Construction of the Regions}\label{sec:smc}
Our numerical algorithm fits within the subclass of Monte Carlo methods called \emph{sequential Monte Carlo} or SMC\footnote{See \cite{Granade2012Robust} and references therein for a more detailed description of the algorithm and \cite{Ferrie2012Qinfer} for an implementation in software.}.  We approximate the posterior distribution by a weighted sum of delta-functions:
\begin{equation}\label{eq:SMC approximation}
\Pr(\vec{x}|D;C) \approx \sum_{j=1}^n w_j(D;C) \delta(\vec{x} - \vec{x}_j),
\end{equation}
where the weights at each step are iteratively calculated from the previous step via
\begin{equation}
w_j(d_{l+1};c_{l+1}) \propto \Pr(d_{l+1}|\vec{x}_j;c_{l+1})w_j(d_l;c_l),
\end{equation}
followed by a normalization step.  The elements of the set $\{\vec{x}_j\}_{j=0}^n$ are called \emph{particles} and are initially chosen by sampling the prior distribution $\Pr(\vec{x})$ and setting each weight $w_j = 1/n$.  In the equations above, $n$ is the number of particles and controls the accuracy of the approximation.  

Note that the approximation in equation \eqref{eq:SMC approximation} is not a particularly good one \emph{per se} (we are approximating a continuous function by a discrete one after all).  However, it does allow us to calculate some quantities of interest with arbitrary accuracy.  Like all Monte Carlo algorithms, it was designed to approximate expectation values, such that
\begin{equation}\label{eq:SMC expectation approximation}
\mathbb{E}_{\vec{x}|D;C}[f(\vec{x})] \approx \sum_{j=1}^n w_j(D;C) f(\vec{x}_j).
\end{equation}
In other words, it allows us to efficiently evaluate difficult multidimensional integrals with respect to the measure defined by the posterior distribution.

The SMC approximation provides a nearly trivial method to approximate HPD credible regions, which surprisingly has been overlooked.  Since the SMC approximate distribution \eqref{eq:SMC approximation} is a discrete distribution, the credible regions will be (at least initially) discrete sets.  In particular, the HPD $\alpha$-credible set, $\hat X_{\rm SMC}$, is defined by the following construction:
\begin{enumerate}
\item Sort the particles according to their weights: $\{\vec{x}_j : w_i \geq w_l \text{ for } i <  l\}$;
\item Collect particles (starting with the highest weighted) until the sum of the collected particle weights is at least $1-\alpha$.  The resulting set is  $\hat X_{\rm SMC}$.
\end{enumerate}
The proof that this an HPD $\alpha$-credible set is as follows.  Assuming the particles are sorted as above.  Begin with the highest weighted particle $\vec{x}_1$ with weight $w_1$.  Then, the set $X_{1} = \vec{x}_1$ clearly has weight $w_1$ and the largest $k$ satisfying equation \eqref{eq:HPD}, in the definition of HPD credible regions, is $k=w_1$.  Now take the set $X_{1,2} = \{\vec{x}_1,\vec{x}_2\}$ with weight $w_1+w_2$.  The largest $k$ is now $k=w_2$.  Iterate this process until we reach the first weight $w_{n_\alpha}$ such that set $X_{1,2,\ldots,n_\alpha} = \{\vec{x}_1,\vec{x}_2,\ldots,\vec{x}_{n_\alpha}\}$ satisfies $\sum_{j=1}^{n_\alpha} w_j \geq 1-\alpha$.  This set will have largest $k= w_{n_\alpha}$.  The set  $X_{1,2,\ldots,n_\alpha}$ is clearly an $\alpha$-credible set but it is also an HPD $\alpha$-credible set since any $k> w_{n_\alpha}$ will result in a set excluding all particles $\vec{x}_{j}$ with $j\geq n_{\alpha}$ and necessarily have weight less than $1-\alpha$.

The immediate problem with $\hat X_{\rm SMC}$ is that it is a discrete set of points and while it is HPD $\alpha$-credible set for the SMC approximated distribution, \emph{any} discrete set has zero measure according the true posterior $\Pr(\vec{x}|D;C)$.  The resolution is quite simple.  Suppose we have some region $\hat X$ which contains $\hat X_{\rm SMC}$.  Then, according to the SMC approximation \eqref{eq:SMC expectation approximation}, 
\begin{equation}
\Pr(\vec{x} \in \hat X) = \mathbb E_x[\mathbbm 1_{\hat X}(x)]\approx \sum_j w_j\mathbbm 1_{\hat X}(\vec{x}_j) \geq 1- \alpha,
\end{equation}
since $\mathbbm 1_{\hat X}(\vec{x}_j)  = 1$ for all $\vec{x}_j$ in $\hat X_{\rm SMC}$.  Thus, any region enclosing the points in $\hat X_{\rm SMC}$ will be a $\alpha$-credible region.

But we do not want just any $\alpha$-credible region.  The HPD requirement is conceptually similar to asking for the region to be as small as possible while maintaining $1-\alpha$ weight.  If we assume (relaxed later on) the credible regions are convex, then we seek the smallest convex set containing $\hat X_{\rm SMC}$.  This defines the \emph{convex hull} of $\hat X_{\rm SMC}$:
\begin{equation}
\hat X_{\rm hull} = {\rm hull}(\hat X_{\rm SMC}).
\end{equation}
Since $\hat X_{\rm hull}$ is a convex polytope in $\mathbb R^{d^2-1}$, it can be most compactly represented by a list of its vertices, which in the absolute worst cases is the entire set of SMC particles.  That is, we require $n(d^2-1)$ numbers to specify $\hat X_{\rm hull}$.  Although certain classes of convex polytopes contain many symmetries and are easy to conceptualize geometrically, specifying the vertices of the convex hull of a random set of points is not most convenient representation for credible regions.

The most efficient way to describe this hull is the smallest ball containing it since this would be described by a location and single radial parameter.  However, a ball would not account for large covariances in the posterior distribution.  To account for these covariances, we will use ellipsoidal regions where $\vec{c}$ and $\mathbf{A}$ define an ellipsoid via the set of states satisfying $(\vec{x}-\vec{c})^{\rm T} \mathbf{A}^{-1}(\vec{x}-\vec{c})\leq1$.   In other words, $\vec{c}$ is the center of the ellipsoid and $\mathbf{A}$ its covariance matrix.  Crucially, we want the smallest ellipse containing the hull, the so-called \emph{minimum volume enclosing ellipse} (MVEE): 
\begin{equation}\label{eq:mvee}
(\vec{c},\mathbf{A}):=\hat X_{\rm MVEE} = {\rm MVEE}(\hat X_{\rm hull}).
\end{equation}
 To numerically construct the MVEE, we use the algorithm of Khachiyan \cite{Todd2007Khachiyans}.

To summarize, $\hat X_{\rm MVEE}$ is the numerical approximation to the HPD credible region.  The posterior covariance regions, $\hat X_{\rm PCE}$, defined earlier in equation \eqref{eq:PCE}, are far less computationally intensive to construct than $\hat X_{\rm MVEE}$ and are expected to be HPD credible regions in the asymptotic limit.  In order to show that they are also approximately HPD credible regions for finite data, we compare $\hat X_{\rm MVEE}$ and $\hat X_{\rm PCE}$ in a number of examples and also look at the performance in cases where limited computational resources prohibit constructing $\hat X_{\rm MVEE}$ over many simulations.

Finally, we note that to compare the sizes of various ellipsoids, we will use the volume
\begin{equation}\label{eq:ellipse vol}
 {\rm Vol}(\mathbf{A}) = \frac{\pi^{(d-1)/2}}{\Gamma(\frac {d}2 + 1)} \det(\mathbf{A})^{-1/2},
\end{equation}
where $d$ is the dimension of the parameter space and $\Gamma(n) = (n-1)!$ is the well-known Gamma function.
 
\section{Examples}\label{sec:ex}

Consider repeated preparations of a qubit subjected to random Pauli measurements.  We label the Pauli operators $\{\sigma_0,\sigma_1,\sigma_2,\sigma_3\}$ such that an arbitrary operator can be written
\begin{equation}
\rho = \frac12(\sigma_0 + x_1 \sigma_1+ x_2 \sigma_2 + x_3 \sigma_3),
\end{equation}
and 
\begin{equation}\label{eq:qubit param}
x_k = \Tr(\rho \sigma_k).
\end{equation}
For many qubits, the situation is similar.  The reconstruction is given by
\begin{equation}
\rho = \frac1{2^n}\sum_k x_k \sigma_k,
\end{equation}
where 
\begin{equation}
\sigma_k = \sigma_{k_1}\otimes\sigma_{k_2}\otimes\cdots\otimes\sigma_{k_n},
\end{equation}
and we index by $k = k_1+ 4 k_2 + 4^2 k_3 + \cdots + 4^{n-1} k_n$.  Then the parameterization is equivalent to that in equation \eqref{eq:qubit param}.  Since each Pauli is idempotent, $\sigma_k^2= \mathbbm 1$, each individual measurement has $2$ possible outcomes which we label $d \in \{+1,-1\}$ for the $+1$ and $-1$ eigenvalues.  The likelihood function of a single measurement is then
\begin{equation}
\Pr(\pm 1|\vec{x},\sigma_k) = \frac{1 \pm x_k}{2}.
\end{equation}
We also consider the effect of errors.  We will not assume a particular model for the errors since any error model, for our two outcome measurements, manifests as a bit flip, or equivalently, randomization channel.  For simplicity we assume the process is symmetric so we have a single parameter $\eta$, called the \emph{visibility}, which has the following effect on the likelihood function:
\begin{equation}
\Pr(\pm 1|\vec{x},\eta, \sigma_k) = \eta \frac{1 \pm x_k}{2} + \frac{1-\eta}{2}.
\end{equation}
We consider two cases: the visibility in known and fixed or the visibility is unknown but still fixed run-to-run.  In the former case, the task is to compute the PCE $\hat X_{\rm PCE}$ for the state only, while in the latter case the task is to compute the PCE over the \emph{joint distribution} of $(\vec{x},\eta)$.  If only a region of states is desired, we can orthogonally project the PCE onto the subspace of the parameter space defining the state (and similarly for $\eta$).  Hence, we will have separate marginal PCEs for the state and visibility separately.  

\begin{figure}\centering
  \includegraphics[width=.90\columnwidth]{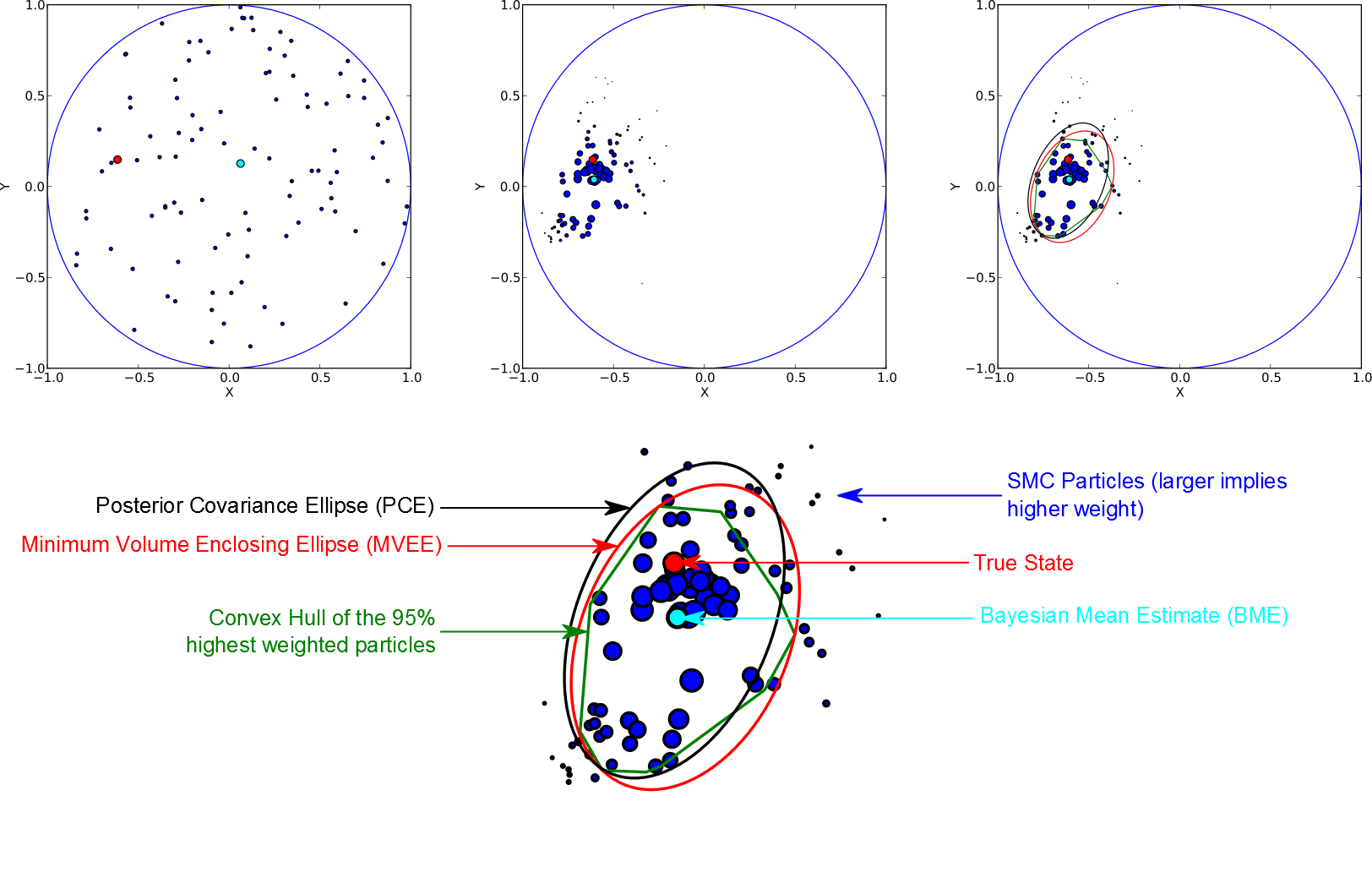}
  \caption{\label{fig:eq_rebit} Construction of regions for a rebit subjected to 100 randomly selected $X$ and $Y$.  On the left, we have the initial 100 particles (blue) randomly select according to the Hilbert-Schmidt prior and the randomly generated ``true'' state (red).  In the middle figure, we have the posterior SMC particle cloud after 100 randomly selected $X$ and $Y$ measurements and the estimated state, the mean of distribution, is shown in teal.  The larger weighted particles are represented as larger dots.  On the right, the same distribution of particles is presented along with the regions discussed in the text.  The green line is the convex hull of those highest weighted particles comprising at least 95\% of the particle weight (this is $\hat X_{\rm hull}$).  The red ellipse is $\hat X_{\rm MVEE}$, the smallest ellipse containing  $\hat X_{\rm hull}$.  Finally, in black, is the ellipse defined by the estimated convariance matrix of the particle cloud, $\hat X_{\rm PCE}$.  These objects are blown up below the figure to show details.}
\end{figure}

\begin{figure}\centering
  \includegraphics[width=.90\columnwidth]{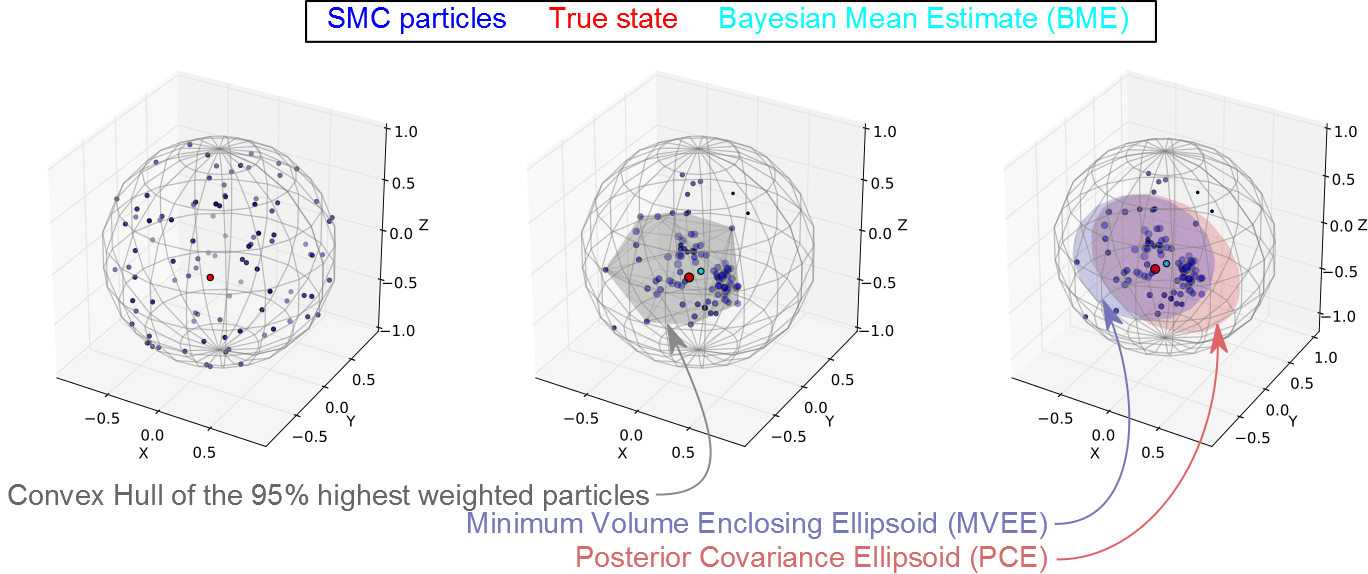}
  \caption{\label{fig:eq_qubit}  Construction of regions for qubit subjected to 20 randomly selected Pauli measurements using the SMC approximation with 100 particles.  On the left, we have the initial 100 particles (blue) randomly select according to the Hilbert-Schmidt prior and the randomly generated ``true'' state (red).  In the middle and right figure, we have the posterior SMC particle cloud after 20 randomly selected Pauli measurements and the estimated state, the mean of distribution, is shown in teal.  The larger weighted particles are represented as larger dots.  In the middle figure, the gray object is the convex hull of those highest weighted particles comprising at least 95\% of the particle weight (this is $\hat X_{\rm hull}$).  In the right figure, the blue ellipsoid is $\hat X_{\rm MVEE}$, the smallest ellipse containing  $\hat X_{\rm hull}$ while the red ellipsoid is the posterior covariance ellipsoid $\hat X_{\rm PCE}$.}
\end{figure}

Examples of how the regions are constructed are presented in figures \ref{fig:eq_rebit} and \ref{fig:eq_qubit}.

\section{Results}\label{sec:results}

\begin{figure}\centering
  \includegraphics[width=.75\columnwidth]{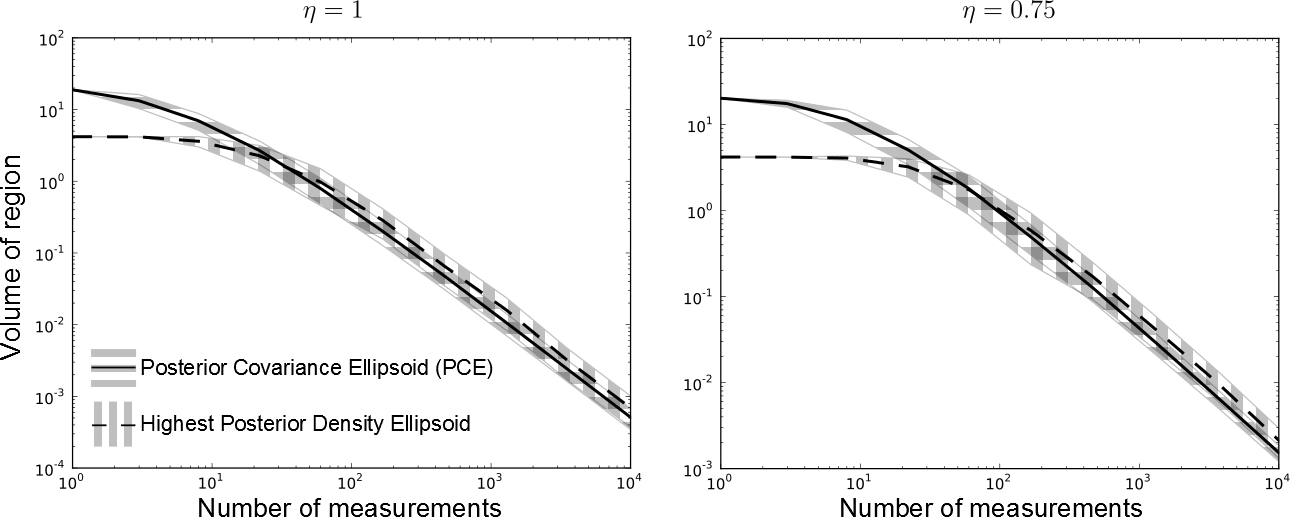}
  \caption{\label{fig:qubit_sizes_known_vis}  The volume of the constructed ellipsoid for (left) perfect measurements and (right) limited visibility measurements.  For all constructed regions, $n = 100,000$ particles where used.  The results are from 100 simulations, where the line represents the mean and the shaded areas are those volumes one standard deviation from the mean.}
\end{figure}

\begin{figure}\centering
  \includegraphics[width=.40\columnwidth]{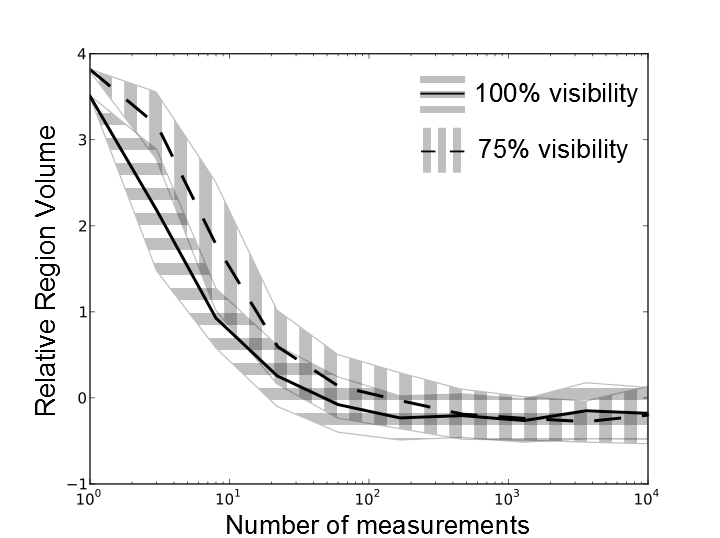}
  \caption{\label{fig:qubit_size_known_vis}  The relative size $|{\rm Vol}(\hat X_{\rm PCE})-{\rm Vol}(\hat X_{\rm MVEE})|/{\rm Vol}(\hat X_{\rm MVEE})$ of the 95\% credible regions, where $\hat X_{\rm MVEE}$ is a numerical approximation to the HPD 95\% credible region.  For both strategies, $n = 100,000$ particles where used.  The results are from 100 simulations, where the line represents the mean and the shaded area are ratios one standard deviation from the mean.  Note that the posterior covariance ellipsoid is on average 10\% smaller than the HPD region after about 100 measurements.}
\end{figure}

\begin{figure}\centering
  \includegraphics[width=.75\columnwidth]{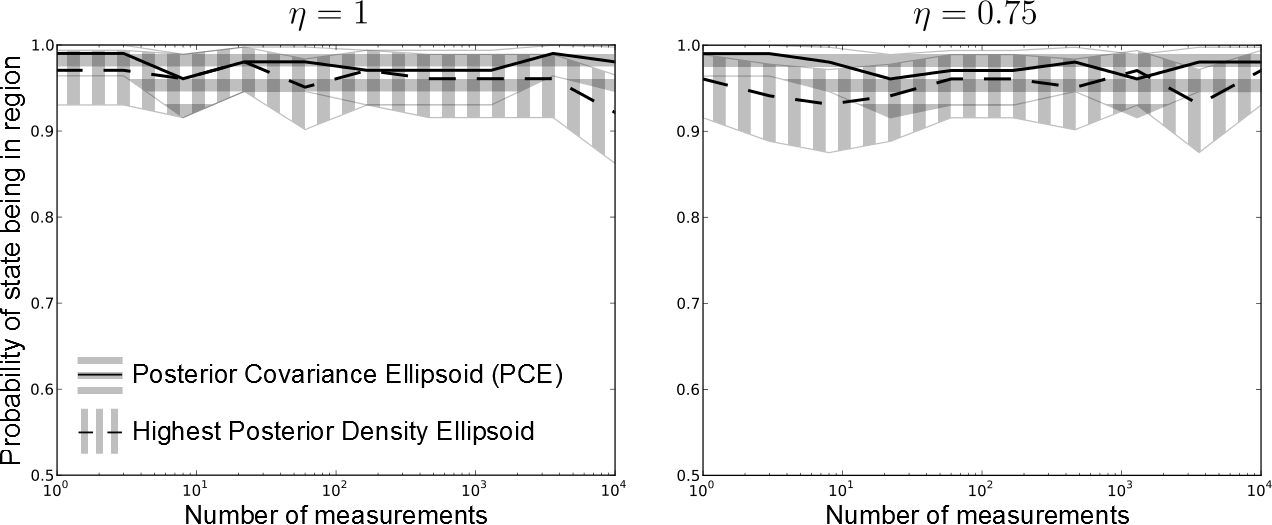}
  \caption{\label{fig:qubit_pr_known_vis}  The probability of the state lying in the constructed ellipsoid for (left) perfect measurements and (right) limited visibility measurements.  In all cases, the target was a 95\% credible region.  For all constructed regions, $n = 100,000$ particles where used.  The results are from 100 simulations, where the line represents the mean and the shaded area is (just to be meta) the HPD 95\% credible region of the probability (derived from the beta distribution and a uniform prior).}
\end{figure}

\begin{figure}\centering
  \includegraphics[width=.4\columnwidth]{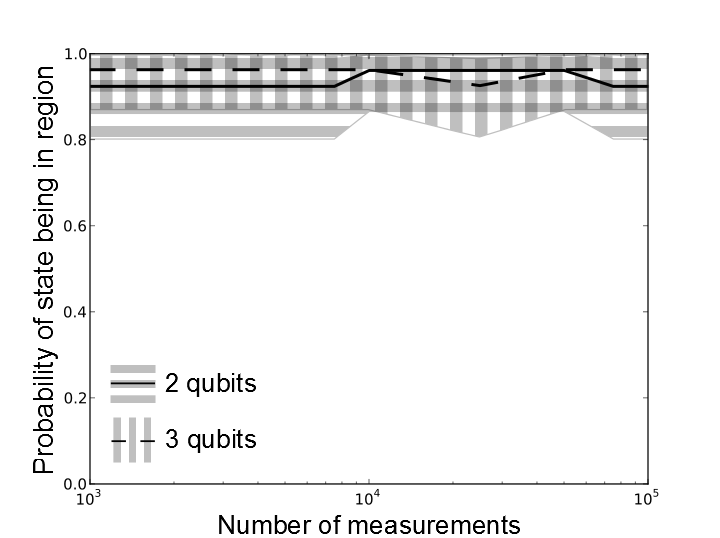}
  \caption{\label{fig:23}  The probability of the state lying in the constructed posterior covariance ellipsoid (approximating the 95\% HPD region) for the two and three qubit model described in the text and using a visibility parameter $\eta = 0.9$.  For all constructed regions, $n = 100,000$ particles where used.  The results are from 25 simulations, where the line represents the mean and the shaded area is the HPD 95\% credible region of the probability (derived from the beta distribution and a uniform prior).}
\end{figure}

We first look at a comparison of $\hat X_{\rm MVEE}$ and $\hat X_{\rm PCE}$.  These results are presented in figures \ref{fig:qubit_sizes_known_vis},\ref{fig:qubit_size_known_vis} and \ref{fig:qubit_pr_known_vis}.  In figures \ref{fig:qubit_sizes_known_vis} and  \ref{fig:qubit_size_known_vis}, the size of the two classes of regions is compared.  Initially the volume of the PCE is not smaller than that of the entire parameter space, which is to be expected since it is motivated from asymptotic normality.  However, it rapidly converges in volume to, and becomes slightly smaller than, $\hat X_{\rm MVEE}$.  Both sets of regions decrease in size at the same rate as a function of the number of measurements.  This suggests that the PCEs are approximate HPD credible regions.  This is important because, as opposed to all other region estimators, PCE regions are computationally tractable.  That the PCEs remain valid in higher dimensions is shown in figure \ref{fig:23}.  In figure \ref{fig:23} the probability for the state to lie in the constructed PCE region is shown to be consistent with the target of 95\% containment probability for two and three-qubits subjected to random Pauli measurements.

\begin{figure}\centering
  \includegraphics[width=.75\columnwidth]{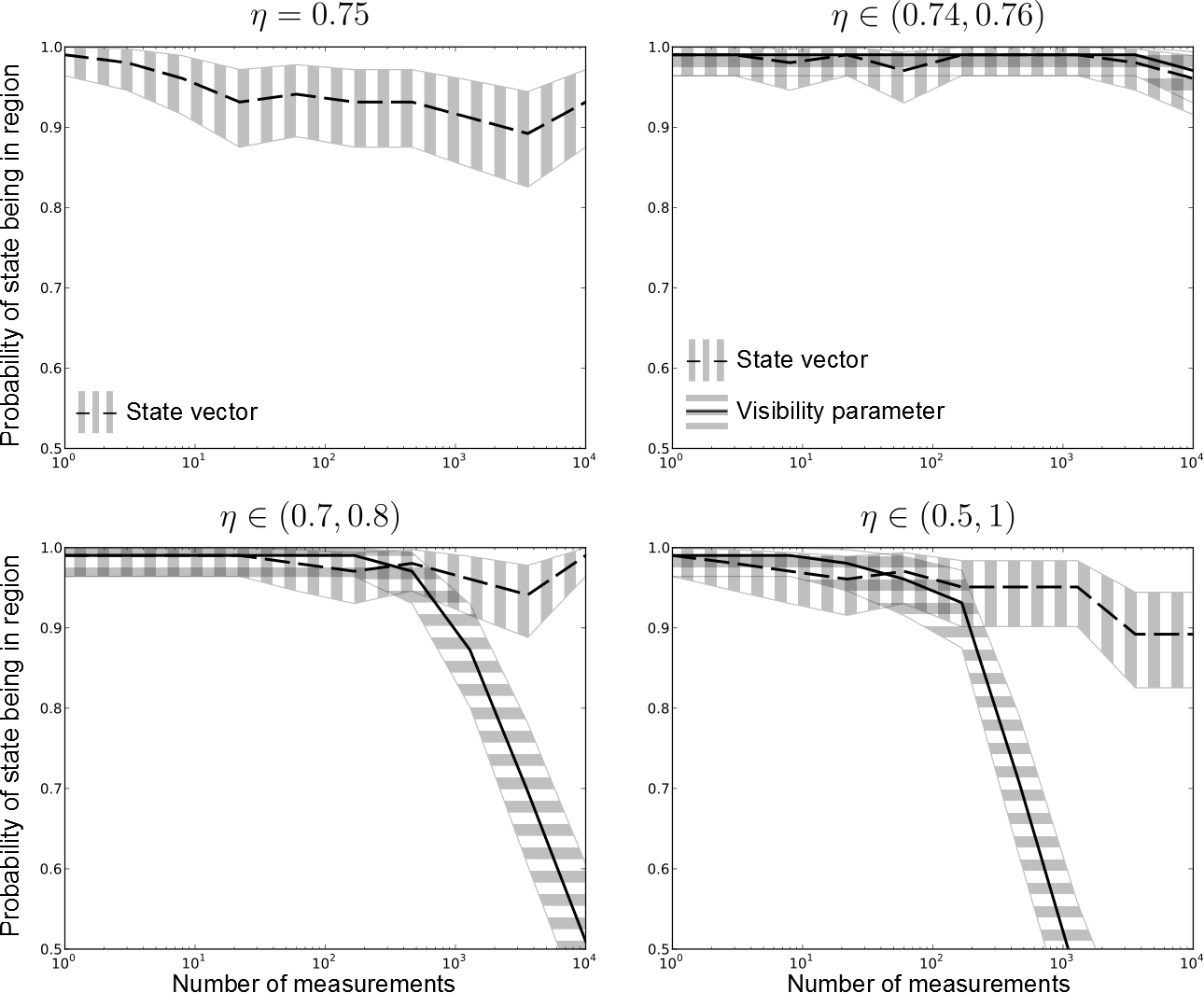}
  \caption{\label{fig:qubit_pr_unknown_vis}  The probability of the state lying in the constructed  (95\% credible) posterior covariance ellipse for varying levels of knowledge of the visibility parameter.  In the upper left, the visibility is known (this is identical to figure \ref{fig:qubit_pr_known_vis}).  In all other figures, the visibility in unknown (but known to lie in the specified interval---with a uniform prior).  For both the state and the visibility parameter, a marginal posterior covariance ellipse is constructed and the tested against the true state and parameter.}
\end{figure}

In the above mentioned cases, the visibility $\eta$ was assumed to be known.  In figure \ref{fig:qubit_pr_unknown_vis}, the case of unknown visibility is considered.  When the visibility in known relatively accurately, the PCE captures the state and visibility accurately.  However, as the initial variance in the prior on the visibility increases, the ability of the PCE to capture the both the state and visibility diminishes.  Surprisingly, the PCE still finds the state even when it cannot resolve the visibility accurately.

\begin{figure}\centering
  \includegraphics[width=.75\columnwidth]{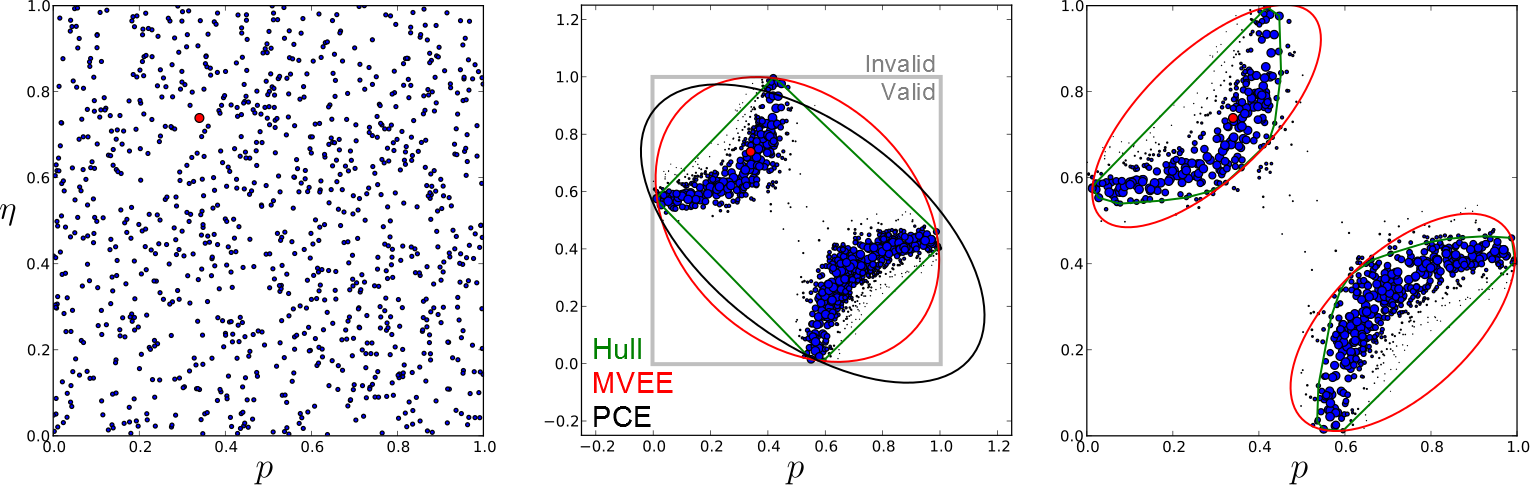}
  \caption{\label{fig:nc} Construction of regions for qubit (known to be of the form $\rho = p|0\rangle\!\langle 0| +(1-p)|0\rangle\!\langle 0|$) with unknown visibility subjected to 1000 randomly selected $Z$ measurements.  On the left, we have the initial 1000 particles (blue) randomly select according to a uniform prior and the randomly generated ``true'' state (red).  In the middle figure, we have the posterior SMC particle cloud after 1000 randomly selected $Z$ measurements along with the following regions: the green line is the convex hull of those highest weighted particles comprising at least 95\% of the particle weight (this is $\hat X_{\rm hull}$);  the red ellipse is $\hat X_{\rm MVEE}$, the smallest ellipse containing  $\hat X_{\rm hull}$; and, in black is the ellipse defined by the estimated convariance matrix of the particle cloud, $\hat X_{\rm PCE}$.  When the posterior is disjoint, all regions poorly approximate the HPD credible region.  On the right, the same distribution of particles is shown along with the convex hull and MVEE regions after the modes of the distribution have been identified via the DBSCAN clustering algorithm. }
\end{figure}

The problem is easily identified to be the assumption that this posterior has a single mode with a convex HPD credible region.  To illustrate the problem graphically, we need to reduce the dimensionality.  To this end, we assume the state is of the form $\rho = p|0\rangle\!\langle 0| +(1-p)|0\rangle\!\langle 0|$, with $p$ unknown and to be estimated along with the visibility $\eta$.  A typical example of a posterior distribution and the possible regions is shown in figure \ref{fig:nc}.  There are two things to note: (1) the posterior distribution has two modes; (2) even within each mode, the distribution is not well approximated by a Gaussian.  Both of these are due to the degeneracy in the posterior distribution of $(p,\eta)$ arising from the symmetry in of $p$ and $\eta$ in the likelihood function.  For example, an outcome could equally well be explained by a large $p$ and small $\eta$ as a small $\eta$ and large $p$.

The problem of many modes in the posterior can be resolved by reporting disjoint ellipsoidal regions, one for each mode.  The discrete set of highest weighted SMC points $X_{\rm SMC}$ naturally find themselves within the modes.  Given this set, the task is to then identify which particles belong to which modes.  In machine learning parlance, this is the problem of \emph{clustering}.  Many solutions to this problem exist, each with its own set of advantages and drawbacks.  Here we have used DBSCAN \cite{Ester1996Densitybased,Pedregosa2011Scikitlearn}, as it seems to require the fewest number of assumptions\footnote{This clustering algorithm was suggested to the author by Chris Granade (as noted in Ref. \cite{Granade2012Robust}).}.

\begin{figure}\centering
  \includegraphics[width=.85\columnwidth]{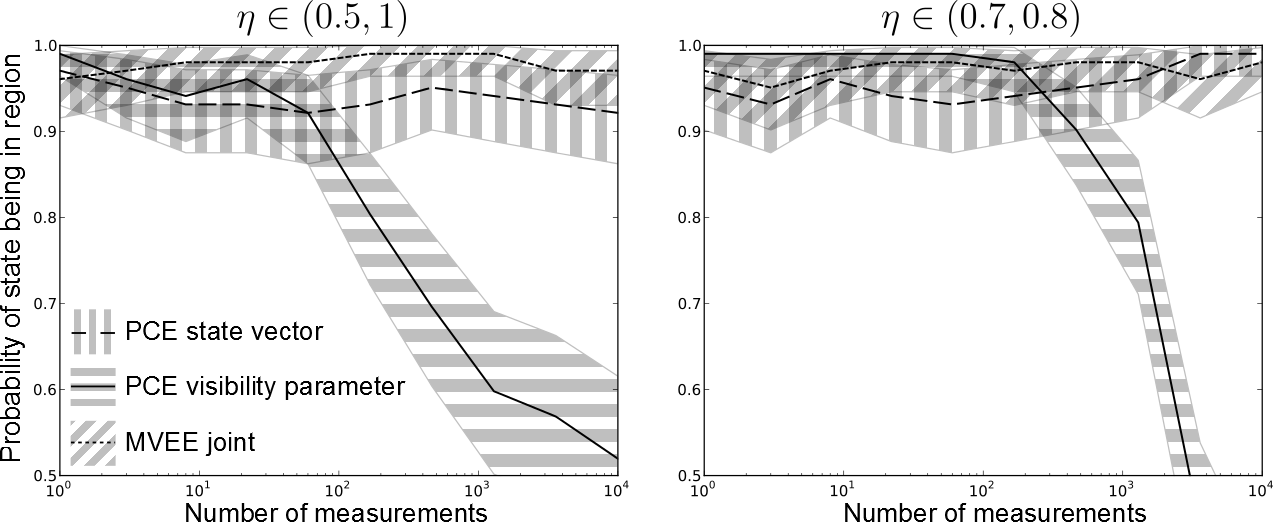}
  \caption{\label{fig:mveevis}   The probability of the state lying in the (95\% credible) posterior covariance ellipse $\hat X_{\rm PCE}$ or the minimum volume enclosing ellipse $\hat X_{\rm MVEE}$ for varying levels of knowledge of the visibility parameter.  The visibility in unknown (but known to lie in the specified interval---with a uniform prior).  For both the state and the visibility parameter, a marginal posterior covariance ellipse is constructed and the tested against the true state and parameter.}
\end{figure}

When the visibility is known fairly well only the first problem, disjoint regions, is automatically resolved.  In other words, it is not likely that the PCE will be the optimal region estimator unless the visibility is relatively well-known.  Practically, when the noise is known with some---but not perfect---accuracy, the MVEE region $\hat X_{\rm MVEE}$ still behaves properly even when the PCE region $\hat X_{\rm PCE}$ does not.  This is demonstrated in figure \ref{fig:mveevis} where we see that $\hat X_{\rm MVEE}$ contains the true state with the correct probability but $\hat X_{\rm PCE}$ does not.  In practice then, the recommendation is to identify whether the problem specifies convex credible regions.  If so, then the PCE is the appropriate choice; if not, then a clustering algorithm should be used to identify the modes of the distribution first.

\section{Conclusion}\label{sec:end}

For the three qubit data shown in figure \ref{fig:23}, the constructed PCEs were ellipsoids in a $4^3-1=63$ dimensional parameters space.  That these simulations were performed on an average laptop computer lends credence to the claim that the regions discussed here, constructed via the SMC algorithm, are a computationally attractive solution to the problem of providing region estimators of quantum states.  For the implementation given in \cite{Ferrie2012Qinfer}, further optimization of the code itself would be required to move beyond 100-or-so parameters for conventional computing resources.  Alternative to this is the ``throw money at it'' solution and run the code on high performance computing machines.

The key obstacle to a fully scalable solution is the \emph{curse of dimensionality}, which is ever-present in quantum theory.   The problem that the parameter space grows exponentially presents a challenge in representing the quantum state and simulating the dynamics of the quantum system.  Both of these obstacles can be overcome within the computational framework presented here.  To address these problems, we follow the already many ingenious methods reducing the complexity
of identifying and characterizing quantum states and processes.  These include identifying stabilizer states \cite{Gottesman2008Identifying}; tomography for matrix product states \cite{Cramer2010Efficient}; tomography for permutationally invariant states \cite{Toth2010Permutationally}; learning local Hamiltonians \cite{daSilva2011Practical};
tomography for low-rank states via compressed sensing \cite{Flammia2012Quantum}; and
tomography for multi-scale entangled states
\cite{LandonCardinal2012Practical}.  These techniques employ efficient simulation algorithms which propagate efficient representations of the state vector to calculate of the probabilities defining the likelihood function.  Physical constraints and careful engineering lead to such drastic dimensional reductions that we can use along with additional prior information in the Bayesian method.  

The above mentioned methods aimed at reducing the complexity of estimating parameters has relied on the notion of \emph{strong} simulation, where the likelihood function is computed exactly.  On the other hand is \emph{weak} simulation, where the likelihood is not computed but is instead sampled from.  The distinction between strong and weak simulation has been a topic of recent interest in quantum computational complexity where it has been shown, for example, that there is exists subtheories of quantum mechanics which admit efficient weak simulation but do not allow for efficient strong simulation \cite{Aaronson2010Computational,VandenNest2011Simulating}.  It has recently been shown that the Bayesian sequential Monte Carlo algorithm can also be used in the case where one only has access to a weak simulator \cite{Ferrie2013Likelihoodfree}.  

Finally, we might find ourselves with a quantum system which does not admit efficient strong nor weak simulation.  In that case, it still may be possible to efficiently characterize the system using a trusted \emph{quantum} simulator.  For the case of estimating dynamical parameters, it has been shown that the Bayesian SMC approach can also perform estimation tasks efficiently using a quantum simulator \cite{Wiebe2013Hamiltonian}.

In this work we have considered supplementing point estimates of quantum states with \emph{regions} of state space.  These regions contain the true state with a pre-determined probability and within the tolerance of the numerical algorithm.  The numerical algorithm has tunable parameters which trades accuracy for computational efficiency and thus can be determined based on desired optimality and available computational resources.  When the noise is known with relatively high accuracy, the optimal regions are the \emph{posterior covariance ellipsoids}.  When the noise is unknown, more complex techniques are available to construct ellipsoids which capture the state.  In any case, the constructed regions are ellipsoids which are easily described and conceptualized.

In the context of classical statistics, quantum state estimation can simply be thought of as overly ambitious parameter estimation.  That is, quantum state estimation is just classical parameter estimation with a specific model and, perhaps, oddly appearing constraints.  The point is that the framework presented here for region estimation is suitable to any parameter estimation problem.  In particular, we have already shown that additional noise on the measurements can be estimated \emph{simultaneously} with the unknown state.  More generally, the framework possesses a beautiful modularity which allows arbitrary statistical models to be learned.

\begin{acknowledgements}
The algorithm presented here has been implemented in a software package called \emph{Qinfer} \cite{Ferrie2012Qinfer} using the Python programming language and the scientific computation library \emph{SciPy} \cite{Jones2001SciPy}.  The author thanks Chris Granade for many discussions, timely advice and the majority of contributions necessary to make this open-source software a reality, which has indeed proven useful in many of our collaborations.  This work was supported by the National Science Foundation Grant Nos.~PHY-1212445 and~PHY-1314763, by Office of Naval Research Grant No.~N00014-11-1-0082, and by the Canadian Government through the NSERC PDF program.
\end{acknowledgements}

\bibliographystyle{apsrev}

\begin{thebibliography}{32}
\expandafter\ifx\csname natexlab\endcsname\relax\def\natexlab#1{#1}\fi
\expandafter\ifx\csname bibnamefont\endcsname\relax
  \def\bibnamefont#1{#1}\fi
\expandafter\ifx\csname bibfnamefont\endcsname\relax
  \def\bibfnamefont#1{#1}\fi
\expandafter\ifx\csname citenamefont\endcsname\relax
  \def\citenamefont#1{#1}\fi
\expandafter\ifx\csname url\endcsname\relax
  \def\url#1{\texttt{#1}}\fi
\expandafter\ifx\csname urlprefix\endcsname\relax\def\urlprefix{URL }\fi
\providecommand{\bibinfo}[2]{#2}
\providecommand{\eprint}[2][]{\url{#2}}

\bibitem[{\citenamefont{Blume-Kohout}(2010)}]{BlumeKohout2010Optimal}
\bibinfo{author}{\bibfnamefont{R.}~\bibnamefont{Blume-Kohout}},
  \bibinfo{journal}{New Journal of Physics} \textbf{\bibinfo{volume}{12}},
  \bibinfo{pages}{043034} (\bibinfo{year}{2010}), 
  \urlprefix\url{http://dx.doi.org/10.1088/1367-2630/12/4/043034}.

\bibitem[{\citenamefont{Home et~al.}(2009)\citenamefont{Home, Hanneke, Jost,
  Amini, Leibfried, and Wineland}}]{Home2009Complete}
\bibinfo{author}{\bibfnamefont{J.~P.} \bibnamefont{Home}},
  \bibinfo{author}{\bibfnamefont{D.}~\bibnamefont{Hanneke}},
  \bibinfo{author}{\bibfnamefont{J.~D.} \bibnamefont{Jost}},
  \bibinfo{author}{\bibfnamefont{J.~M.} \bibnamefont{Amini}},
  \bibinfo{author}{\bibfnamefont{D.}~\bibnamefont{Leibfried}},
  \bibnamefont{and} \bibinfo{author}{\bibfnamefont{D.~J.}
  \bibnamefont{Wineland}}, \bibinfo{journal}{Science}
  \textbf{\bibinfo{volume}{325}}, \bibinfo{pages}{1227} (\bibinfo{year}{2009}),
  \urlprefix\url{http://dx.doi.org/10.1126/science.1177077}.

\bibitem[{\citenamefont{Blume-Kohout}(2012)}]{BlumeKohout2012Robust}
\bibinfo{author}{\bibfnamefont{R.}~\bibnamefont{Blume-Kohout}},
  \emph{\bibinfo{title}{{Robust error bars for quantum tomography}}}
  (\bibinfo{year}{2012}), \eprint{arXiv:1202.5270},
  \urlprefix\url{http://arxiv.org/abs/1202.5270}.

\bibitem[{\citenamefont{\v{R}eh\'{a}\v{c}ek
  et~al.}(2008)\citenamefont{\v{R}eh\'{a}\v{c}ek, Mogilevtsev, and
  Hradil}}]{Rehacek2008Tomography}
\bibinfo{author}{\bibfnamefont{J.}~\bibnamefont{\v{R}eh\'{a}\v{c}ek}},
  \bibinfo{author}{\bibfnamefont{D.}~\bibnamefont{Mogilevtsev}},
  \bibnamefont{and} \bibinfo{author}{\bibfnamefont{Z.}~\bibnamefont{Hradil}},
  \bibinfo{journal}{New Journal of Physics} \textbf{\bibinfo{volume}{10}},
  \bibinfo{pages}{043022} (\bibinfo{year}{2008}), 
  \urlprefix\url{http://dx.doi.org/10.1088/1367-2630/10/4/043022}.

\bibitem[{\citenamefont{Flammia et~al.}(2012)\citenamefont{Flammia, Gross, Liu,
  and Eisert}}]{Flammia2012Quantum}
\bibinfo{author}{\bibfnamefont{S.~T.} \bibnamefont{Flammia}},
  \bibinfo{author}{\bibfnamefont{D.}~\bibnamefont{Gross}},
  \bibinfo{author}{\bibfnamefont{Y.-K.} \bibnamefont{Liu}}, \bibnamefont{and}
  \bibinfo{author}{\bibfnamefont{J.}~\bibnamefont{Eisert}},
  \bibinfo{journal}{New Journal of Physics} \textbf{\bibinfo{volume}{14}},
  \bibinfo{pages}{095022} (\bibinfo{year}{2012}),
  \urlprefix\url{http://dx.doi.org/10.1088/1367-2630/14/9/095022}.

\bibitem[{\citenamefont{Sugiyama et~al.}(2011)\citenamefont{Sugiyama, Turner,
  and Murao}}]{Sugiyama2011Error}
\bibinfo{author}{\bibfnamefont{T.}~\bibnamefont{Sugiyama}},
  \bibinfo{author}{\bibfnamefont{P.~S.} \bibnamefont{Turner}},
  \bibnamefont{and} \bibinfo{author}{\bibfnamefont{M.}~\bibnamefont{Murao}},
  \bibinfo{journal}{Physical Review A} \textbf{\bibinfo{volume}{83}},
  \bibinfo{pages}{012105} (\bibinfo{year}{2011}),
  \urlprefix\url{http://dx.doi.org/10.1103/physreva.83.012105}.

\bibitem[{\citenamefont{Sugiyama et~al.}(2013)\citenamefont{Sugiyama, Turner,
  and Murao}}]{Sugiyama2013Precisionguaranteed}
\bibinfo{author}{\bibfnamefont{T.}~\bibnamefont{Sugiyama}},
  \bibinfo{author}{\bibfnamefont{P.~S.} \bibnamefont{Turner}},
  \bibnamefont{and} \bibinfo{author}{\bibfnamefont{M.}~\bibnamefont{Murao}},
  \emph{\bibinfo{title}{{Precision-guaranteed quantum tomography}}}
  (\bibinfo{year}{2013}), \eprint{arXiv:1306.4191},
  \urlprefix\url{http://arxiv.org/abs/1306.4191}.

\bibitem[{\citenamefont{Christandl and Renner}(2011)}]{Christandl2011Reliable}
\bibinfo{author}{\bibfnamefont{M.}~\bibnamefont{Christandl}} \bibnamefont{and}
  \bibinfo{author}{\bibfnamefont{R.}~\bibnamefont{Renner}},
Physical Review Letters {\bf 109}, 120403 (2012),
  \urlprefix\url{http://dx.doi.org/10.1103/PhysRevLett.109.120403}.

\bibitem[{\citenamefont{Shang et~al.}(2013)\citenamefont{Shang, Ng, Sehrawat,
  Li, and Englert}}]{Shang2013Optimal}
\bibinfo{author}{\bibfnamefont{J.}~\bibnamefont{Shang}},
  \bibinfo{author}{\bibfnamefont{H.~K.} \bibnamefont{Ng}},
  \bibinfo{author}{\bibfnamefont{A.}~\bibnamefont{Sehrawat}},
  \bibinfo{author}{\bibfnamefont{X.}~\bibnamefont{Li}}, \bibnamefont{and}
  \bibinfo{author}{\bibfnamefont{B.-G.} \bibnamefont{Englert}},
  \emph{\bibinfo{title}{{Optimal error regions for quantum state estimation}}}
  (\bibinfo{year}{2013}), \eprint{arXiv:1302.4081},
  \urlprefix\url{http://arxiv.org/abs/1302.4081}.

\bibitem[{\citenamefont{Husz\'{a}r and Houlsby}(2012)}]{Huszar2012Adaptive}
\bibinfo{author}{\bibfnamefont{F.}~\bibnamefont{Husz\'{a}r}} \bibnamefont{and}
  \bibinfo{author}{\bibfnamefont{N.~M.~T.} \bibnamefont{Houlsby}},
  \bibinfo{journal}{Physical Review A} \textbf{\bibinfo{volume}{85}} 052120
  (\bibinfo{year}{2012}),
  \urlprefix\url{http://dx.doi.org/10.1103/PhysRevA.85.052120}.

\bibitem[{\citenamefont{Chase and Geremia}(2009)}]{Chase2009Singleshot}
\bibinfo{author}{\bibfnamefont{B.~A.} \bibnamefont{Chase}} \bibnamefont{and}
  \bibinfo{author}{\bibfnamefont{J.~M.} \bibnamefont{Geremia}},
  \bibinfo{journal}{Physical Review A} \textbf{\bibinfo{volume}{79}},
  \bibinfo{pages}{022314} (\bibinfo{year}{2009}),
  \urlprefix\url{http://dx.doi.org/10.1103/physreva.79.022314}.

\bibitem[{\citenamefont{Granade et~al.}(2012)\citenamefont{Granade, Ferrie,
  Wiebe, and Cory}}]{Granade2012Robust}
\bibinfo{author}{\bibfnamefont{C.~E.} \bibnamefont{Granade}},
  \bibinfo{author}{\bibfnamefont{C.}~\bibnamefont{Ferrie}},
  \bibinfo{author}{\bibfnamefont{N.}~\bibnamefont{Wiebe}}, \bibnamefont{and}
  \bibinfo{author}{\bibfnamefont{D.~G.} \bibnamefont{Cory}},
  \bibinfo{journal}{New Journal of Physics} \textbf{\bibinfo{volume}{14}},
  \bibinfo{pages}{103013} (\bibinfo{year}{2012}), 
  \urlprefix\url{http://dx.doi.org/10.1088/1367-2630/14/10/103013}.

\bibitem{Wiebe2013Hamiltonian}
\bibinfo{author}{\bibfnamefont{N.}~\bibnamefont{Wiebe}},
\bibinfo{author}{\bibfnamefont{C.~E.} \bibnamefont{Granade}},
  \bibinfo{author}{\bibfnamefont{C.}~\bibnamefont{Ferrie}},
   \bibnamefont{and}
  \bibinfo{author}{\bibfnamefont{D.~G.} \bibnamefont{Cory}},
  \emph{\bibinfo{title}{{Hamiltonian Learning and Certification using Quantum Resources}}}
  (\bibinfo{year}{2013}), \eprint{arXiv:1309.0876},
  \urlprefix\url{http://arxiv.org/abs/1309.0876}.

\bibitem[{\citenamefont{Rosset et~al.}(2012)\citenamefont{Rosset, Sch\"{o}bitz,
  Bancal, Gisin, and Liang}}]{Rosset2012Imperfect}
\bibinfo{author}{\bibfnamefont{D.}~\bibnamefont{Rosset}},
  \bibinfo{author}{\bibfnamefont{R.~F.} \bibnamefont{Sch\"{o}bitz}},
  \bibinfo{author}{\bibfnamefont{J.~D.} \bibnamefont{Bancal}},
  \bibinfo{author}{\bibfnamefont{N.}~\bibnamefont{Gisin}}, \bibnamefont{and}
  \bibinfo{author}{\bibfnamefont{Y.~C.} \bibnamefont{Liang}},
  \bibinfo{journal}{Physical Review A} \textbf{\bibinfo{volume}{86}},
  \bibinfo{pages}{062325} (\bibinfo{year}{2012}),
  \urlprefix\url{http://dx.doi.org/10.1103/physreva.86.062325}.

\bibitem[{\citenamefont{Moroder et~al.}(2013)\citenamefont{Moroder, Kleinmann,
  Schindler, Monz, G\"{u}hne, and Blatt}}]{Moroder2013Certifying}
\bibinfo{author}{\bibfnamefont{T.}~\bibnamefont{Moroder}},
  \bibinfo{author}{\bibfnamefont{M.}~\bibnamefont{Kleinmann}},
  \bibinfo{author}{\bibfnamefont{P.}~\bibnamefont{Schindler}},
  \bibinfo{author}{\bibfnamefont{T.}~\bibnamefont{Monz}},
  \bibinfo{author}{\bibfnamefont{O.}~\bibnamefont{G\"{u}hne}},
  \bibnamefont{and} \bibinfo{author}{\bibfnamefont{R.}~\bibnamefont{Blatt}},
  \bibinfo{journal}{Physical Review Letters} \textbf{\bibinfo{volume}{110}},
  \bibinfo{pages}{180401} (\bibinfo{year}{2013}),
  \urlprefix\url{http://dx.doi.org/10.1103/physrevlett.110.180401}.

\bibitem[{\citenamefont{Langford}(2013)}]{Langford2013Errors}
\bibinfo{author}{\bibfnamefont{N.~K.} \bibnamefont{Langford}},
  \bibinfo{journal}{New Journal of Physics} \textbf{\bibinfo{volume}{15}},
  \bibinfo{pages}{035003} (\bibinfo{year}{2013}),
  \urlprefix\url{http://dx.doi.org/10.1088/1367-2630/15/3/035003}.

\bibitem[{\citenamefont{van Enk and Blume-Kohout}(2013)}]{vanEnk2013When}
\bibinfo{author}{\bibfnamefont{S.~J.} \bibnamefont{van Enk}} \bibnamefont{and}
  \bibinfo{author}{\bibfnamefont{R.}~\bibnamefont{Blume-Kohout}},
  \bibinfo{journal}{New Journal of Physics} \textbf{\bibinfo{volume}{15}},
  \bibinfo{pages}{025024} (\bibinfo{year}{2013}), 
  \urlprefix\url{http://dx.doi.org/10.1088/1367-2630/15/2/025024}.

\bibitem[{\citenamefont{Schwarz and van Enk}(2013)}]{Schwarz2013Error}
\bibinfo{author}{\bibfnamefont{L.}~\bibnamefont{Schwarz}} \bibnamefont{and}
  \bibinfo{author}{\bibfnamefont{S.}~\bibnamefont{van Enk}},
  \emph{\bibinfo{title}{{Error models in quantum computation: an application of
  model selection}}} (\bibinfo{year}{2013}), \eprint{arXiv:1307.0858},
  \urlprefix\url{http://arxiv.org/abs/1307.0858}.

\bibitem[{\citenamefont{Blume-Kohout and
  Hayden}(2006)}]{BlumeKohout2006Accurate}
\bibinfo{author}{\bibfnamefont{R.}~\bibnamefont{Blume-Kohout}}
  \bibnamefont{and} \bibinfo{author}{\bibfnamefont{P.}~\bibnamefont{Hayden}}
  (\bibinfo{year}{2006}), \eprint{arXiv:quant-ph/0603116},
  \urlprefix\url{http://arxiv.org/abs/quant-ph/0603116}.

\bibitem[{\citenamefont{Robert}(2007)}]{Robert2007The}
\bibinfo{author}{\bibfnamefont{C.~P.} \bibnamefont{Robert}},
  \emph{\bibinfo{title}{{The Bayesian Choice: From Decision-Theoretic
  Foundations to Computational Implementation}}}, Springer Texts in Statistics
  (\bibinfo{publisher}{Springer Verlag, New York}, \bibinfo{year}{2007}),
  \bibinfo{edition}{2nd} ed., ISBN \bibinfo{isbn}{0387715983}.

\bibitem[{\citenamefont{Ferrie and Granade}(2012 -)}]{Ferrie2012Qinfer}
\bibinfo{author}{\bibfnamefont{C.}~\bibnamefont{Ferrie}} \bibnamefont{and}
  \bibinfo{author}{\bibfnamefont{C.}~\bibnamefont{Granade}},
  \emph{\bibinfo{title}{{QInfer: Library for statistical inference in quantum
  information}}} (\bibinfo{year}{2012--}),
  \urlprefix\url{http://github.com/csferrie/python-qinfer}.

\bibitem[{\citenamefont{Todd and Y{\i}ld{\i}r{\i}m}(2007)}]{Todd2007Khachiyans}
\bibinfo{author}{\bibfnamefont{M.~J.} \bibnamefont{Todd}} \bibnamefont{and}
  \bibinfo{author}{\bibfnamefont{E.~A.} \bibnamefont{Y{\i}ld{\i}r{\i}m}},
  \bibinfo{journal}{Discrete Applied Mathematics}
  \textbf{\bibinfo{volume}{155}}, \bibinfo{pages}{1731} (\bibinfo{year}{2007}),
  \urlprefix\url{http://dx.doi.org/10.1016/j.dam.2007.02.013}.

\bibitem[{\citenamefont{Ester et~al.}(1996)\citenamefont{Ester, Kriegel, Jorg,
  and Xu}}]{Ester1996Densitybased}
\bibinfo{author}{\bibfnamefont{M.}~\bibnamefont{Ester}},
  \bibinfo{author}{\bibfnamefont{H.-p.} \bibnamefont{Kriegel}},
  \bibinfo{author}{\bibfnamefont{S.}~\bibnamefont{Jorg}}, \bibnamefont{and}
  \bibinfo{author}{\bibfnamefont{X.}~\bibnamefont{Xu}}, in
  \emph{\bibinfo{booktitle}{Proceedings of 2nd International Conference on
  KDD}} (\bibinfo{year}{1996}), pp. \bibinfo{pages}{226--231},
  \urlprefix\url{http://citeseerx.ist.psu.edu/viewdoc/summary?doi=10.1.1.71.1980}.

\bibitem[{\citenamefont{Pedregosa et~al.}(2011)\citenamefont{Pedregosa,
  Varoquaux, Gramfort, Michel, Thirion, Grisel, Blondel, Prettenhofer, Weiss,
  Dubourg et~al.}}]{Pedregosa2011Scikitlearn}
\bibinfo{author}{\bibfnamefont{F.}~\bibnamefont{Pedregosa}},
  \bibinfo{author}{\bibfnamefont{G.}~\bibnamefont{Varoquaux}},
  \bibinfo{author}{\bibfnamefont{A.}~\bibnamefont{Gramfort}},
  \bibinfo{author}{\bibfnamefont{V.}~\bibnamefont{Michel}},
  \bibinfo{author}{\bibfnamefont{B.}~\bibnamefont{Thirion}},
  \bibinfo{author}{\bibfnamefont{O.}~\bibnamefont{Grisel}},
  \bibinfo{author}{\bibfnamefont{M.}~\bibnamefont{Blondel}},
  \bibinfo{author}{\bibfnamefont{P.}~\bibnamefont{Prettenhofer}},
  \bibinfo{author}{\bibfnamefont{R.}~\bibnamefont{Weiss}},
  \bibinfo{author}{\bibfnamefont{V.}~\bibnamefont{Dubourg}},
  \bibnamefont{et~al.}, \bibinfo{journal}{Journal of Machine Learning Research}
  \textbf{\bibinfo{volume}{12}}, \bibinfo{pages}{2825} (\bibinfo{year}{2011}).

\bibitem[{\citenamefont{Gottesman}(2008)}]{Gottesman2008Identifying}
\bibinfo{author}{\bibfnamefont{D.}~\bibnamefont{Gottesman}}
  (\bibinfo{year}{2008}), \urlprefix\url{http://pirsa.org/08080052/}.

\bibitem[{\citenamefont{Cramer et~al.}(2010)\citenamefont{Cramer, Plenio,
  Flammia, Somma, Gross, Bartlett, Landon-Cardinal, Poulin, and
  Liu}}]{Cramer2010Efficient}
\bibinfo{author}{\bibfnamefont{M.}~\bibnamefont{Cramer}},
  \bibinfo{author}{\bibfnamefont{M.~B.} \bibnamefont{Plenio}},
  \bibinfo{author}{\bibfnamefont{S.~T.} \bibnamefont{Flammia}},
  \bibinfo{author}{\bibfnamefont{R.}~\bibnamefont{Somma}},
  \bibinfo{author}{\bibfnamefont{D.}~\bibnamefont{Gross}},
  \bibinfo{author}{\bibfnamefont{S.~D.} \bibnamefont{Bartlett}},
  \bibinfo{author}{\bibfnamefont{O.}~\bibnamefont{Landon-Cardinal}},
  \bibinfo{author}{\bibfnamefont{D.}~\bibnamefont{Poulin}}, \bibnamefont{and}
  \bibinfo{author}{\bibfnamefont{Y.-K.} \bibnamefont{Liu}},
  \bibinfo{journal}{Nat Commun} \textbf{\bibinfo{volume}{1}},
  \bibinfo{pages}{149} (\bibinfo{year}{2010}),
  \urlprefix\url{http://dx.doi.org/10.1038/ncomms1147}.

\bibitem[{\citenamefont{T\'{o}th et~al.}(2010)\citenamefont{T\'{o}th,
  Wieczorek, Gross, Krischek, Schwemmer, and
  Weinfurter}}]{Toth2010Permutationally}
\bibinfo{author}{\bibfnamefont{G.}~\bibnamefont{T\'{o}th}},
  \bibinfo{author}{\bibfnamefont{W.}~\bibnamefont{Wieczorek}},
  \bibinfo{author}{\bibfnamefont{D.}~\bibnamefont{Gross}},
  \bibinfo{author}{\bibfnamefont{R.}~\bibnamefont{Krischek}},
  \bibinfo{author}{\bibfnamefont{C.}~\bibnamefont{Schwemmer}},
  \bibnamefont{and}
  \bibinfo{author}{\bibfnamefont{H.}~\bibnamefont{Weinfurter}},
  \bibinfo{journal}{Physical Review Letters} \textbf{\bibinfo{volume}{105}},
  \bibinfo{pages}{250403} (\bibinfo{year}{2010}),
  \urlprefix\url{http://dx.doi.org/10.1103/physrevlett.105.250403}.

\bibitem[{\citenamefont{da~Silva et~al.}(2011)\citenamefont{da~Silva, Cardinal,
  and Poulin}}]{daSilva2011Practical}
\bibinfo{author}{\bibfnamefont{M.~P.} \bibnamefont{da~Silva}},
  \bibinfo{author}{\bibfnamefont{O.~L.} \bibnamefont{Cardinal}},
  \bibnamefont{and} \bibinfo{author}{\bibfnamefont{D.}~\bibnamefont{Poulin}},
  \bibinfo{journal}{Physical Review Letters} \textbf{\bibinfo{volume}{107}},
  \bibinfo{pages}{210404} (\bibinfo{year}{2011}),
  \urlprefix\url{http://dx.doi.org/10.1103/physrevlett.107.210404}.

\bibitem[{\citenamefont{Landon-Cardinal and
  Poulin}(2012)}]{LandonCardinal2012Practical}
\bibinfo{author}{\bibfnamefont{O.}~\bibnamefont{Landon-Cardinal}}
  \bibnamefont{and} \bibinfo{author}{\bibfnamefont{D.}~\bibnamefont{Poulin}},
  \emph{\bibinfo{title}{{Practical learning method for multi-scale entangled
  states}}} (\bibinfo{year}{2012}), \eprint{arXiv:1204.0792},
  \urlprefix\url{http://arxiv.org/abs/1204.0792}.

\bibitem[{\citenamefont{Aaronson and
  Arkhipov}(2010)}]{Aaronson2010Computational}
\bibinfo{author}{\bibfnamefont{S.}~\bibnamefont{Aaronson}} \bibnamefont{and}
  \bibinfo{author}{\bibfnamefont{A.}~\bibnamefont{Arkhipov}}
  (\bibinfo{year}{2010}), \eprint{arXiv:1011.3245},
  \urlprefix\url{http://arxiv.org/abs/1011.3245}.

\bibitem[{\citenamefont{Van~den Nest}(2011)}]{VandenNest2011Simulating}
\bibinfo{author}{\bibfnamefont{M.}~\bibnamefont{Van~den Nest}},
  \bibinfo{journal}{Quantum Information \& Computation}
  \textbf{\bibinfo{volume}{11}} (\bibinfo{year}{2011}).

\bibitem[{\citenamefont{Ferrie and Granade}(2013)}]{Ferrie2013Likelihoodfree}
\bibinfo{author}{\bibfnamefont{C.}~\bibnamefont{Ferrie}} \bibnamefont{and}
  \bibinfo{author}{\bibfnamefont{C.~E.} \bibnamefont{Granade}},
  \emph{\bibinfo{title}{{Likelihood-free quantum inference: tomography without
  the Born Rule}}} (\bibinfo{year}{2013}), \eprint{arXiv:1304.5828},
  \urlprefix\url{http://arxiv.org/abs/1304.5828}.

\bibitem[{\citenamefont{Jones et~al.}(2001-)\citenamefont{Jones, Oliphant, and
  Peterson}}]{Jones2001SciPy}
\bibinfo{author}{\bibfnamefont{E.}~\bibnamefont{Jones}},
  \bibinfo{author}{\bibfnamefont{T.}~\bibnamefont{Oliphant}}, \bibnamefont{and}
  \bibinfo{author}{\bibfnamefont{P.}~\bibnamefont{Peterson}},
  \emph{\bibinfo{title}{{SciPy: Open source scientific tools for Python}}}
  (\bibinfo{year}{2001--}), \urlprefix\url{http://www.scipy.org/}.

\end{thebibliography}

\end{document}